# Effect of radioactivity decrease. Is there a link with solar flares?

A.G.Parkhomov

Institute for Time Nature Explorations.
Lomonosov Moscow State University, Moscow, Russia.
http: // www.chronos.msu.ru

Results obtained with multichannel installation created for long-term studies of various processes, are collated with the data published by J.H. Jenkins and E.Fischbach, who found a decrease of $^{54}Mn$ radioactivity near the time of series of solar flares between 5 and 17 December 2006. Analysis of the data from our installation in December 2006 has not revealed any deviations from the usual behaviour of the count rates for $^{90}Sr$-$^{90}Y$, $^{60}Co$ and $^{239}Pu$ sources. The same can be said of the data collected during the period of highly powerful solar flares between 19 October and 4 November 2003. Apparent drops in the count rate were detected between 10 and 12 May 2002 while registering the activity of $^{60}Co$ and on 19 and 20 June 2004 for $^{90}Sr$-$^{90}Y$ source. Around the time of these events, no observations of large solar flares were reported. Thus, proposed link between the drop in the rates of radioactive decay and appearance of solar flares could not be confirmed. From obtained outcomes follows, that the radioactivity drop effect, if it really exists, is rather rare, and that the reason calling this effect unequally influences various radioactive sources.

In [1,2] results of experiments are described which detected a drop up to 0.2% from the average value in the count rate for $^{54}Mn$ source in a time period between 12 and 22 December 2006 (Fig.1). Because of a series of solar flares registered between 5 and 17 December, the authors associated the observed decrease in the radioactivity with solar phenomena.

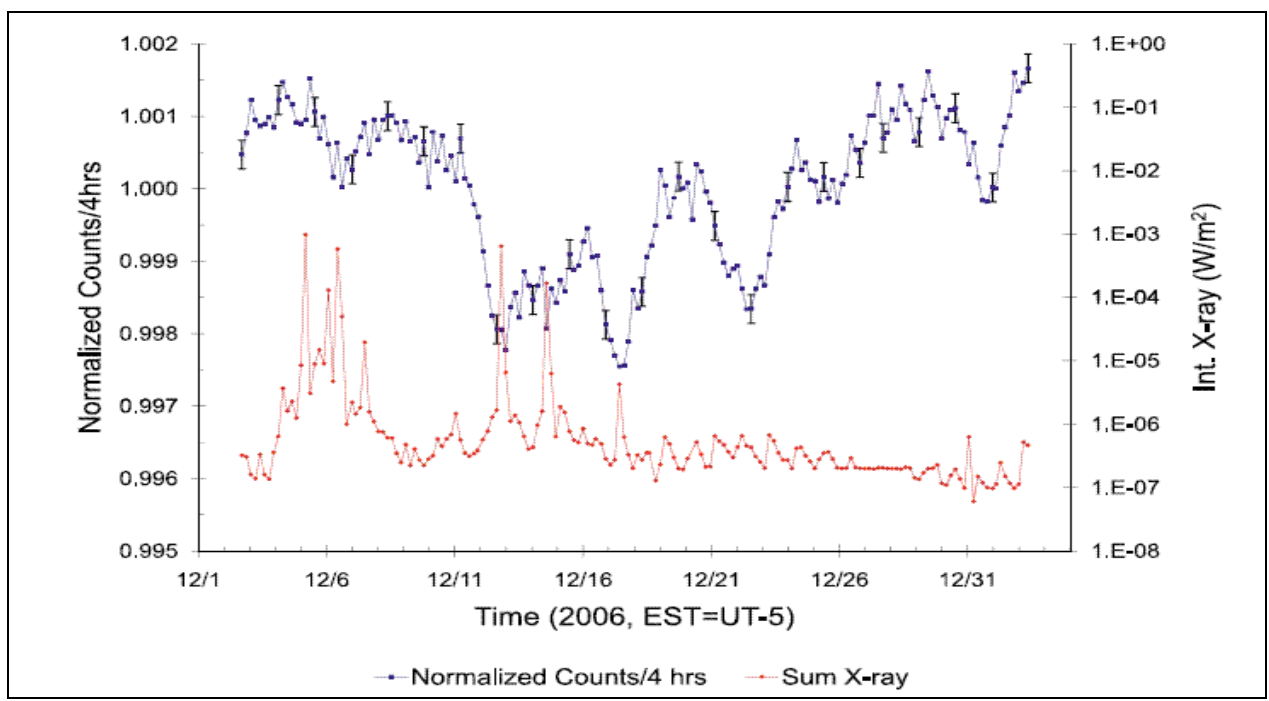

**Fig.1** Normalized December 2006 $^{54}Mn$ decay data along with GOES-11 satellite x-ray data on a logarithmic scale [1]. *NaJ(Tl)* detector was used to register the γ-rays with energy of 834.8 keV, emitted in the decays of $^{54}Mn$. Every point on the $^{54}Mn$ decay curve is a result of a 4-hour recording, divided by the average count rate (relative statistical uncertainty $2 \cdot 10^{-4}$). For the GOES-11 x-ray data, each point is the solar flux in W/m$^2$ summed over the same real-time intervals.

Findings presented in [1,2] are in obvious contradiction with the commonly accepted ideas about radioactivity as a phenomenon immune to external influences. Therefore, it is very important to estimate the extent to which such out of the ordinary studies and their interpretation can be relied upon, by collating them with the results of independent experiments. The results obtained on our installation created for investigations of various processes [3,4], seem to be suitable for such a cross-check. The installation operates almost continuously since 1997, gathering data from 20 channels (such as count rates from α and β sources, electrical oscillations and noise, parameters of the environment).

**Measurement results from December 2006**

The outcomes of radioactivity measurements of $^{90}Sr^{90}Y$ and $^{60}Co$ β sources and $^{239}Pu$ α source are shown in Table 1 and Fig. 2. Highly stable G-M counters were used for registration of β and γ radiation, whereas α particles were detected with silicon detectors, which are practically insensitive to β and γ radiation.

The radiation of $^{90}Sr^{90}Y$ source consists of $^{90}Sr$ β particles (half-life 28.6 years, maximum energy of β particles 546 keV) and β particles of daughter nuclei $^{90}Y$ (half-life 64 hours, maximum energy of β particles 2270 keV). Our installation registers the radiation from these sources simultaneously by two G-M counters СБМ-12 and СТС-5. The counter СБМ-12 reacts to both kinds of β particles. The counter СТС-5 is shielded from the source by 1 mm thick sheet of aluminium and 4.3 mm sheet of polyvinylchloride. It registers β particle from $^{90}Y$ only, let out with energies near the maximal energy of β-spectrum. Sources and counters are located in the thermostat vessel filled with quartz sand.

β decays of $^{60}Co$ (half-life 5.27 years, maximum energy of β particles 314 keV) are accompanied by γ rays with energies 1173 keV and 1332 keV. The radiation of $^{60}Co$ is registered by СБТ-7 counter with a thin micas window passing a majority of β particles radiated from source.

$^{239}Pu$ α source (the half-life 24120 years, energy of α particles 5150 кэВ) is located in the vacuum vessel on a distance 2.5 см from two closely located silicon detectors. Created impulses are registered independently, that allows to separate possible modifications of a radioactivity from noise and instability of an electronics engineering.

Because of low count rate from our installation, for a sufficient decrease of statistical fluctuations, the count rate averaging time was enlarged approximately to 1 day. This is unlikely to affect our ability to discern the effect, since the drop in the $^{54}Mn$ count rate lasted for 12 days.

**Table 1.** Sources, detectors and measurements in December 2006.

| Radionuclides | $^{54}Mn$ | $^{90}Sr^{90}Y$ | $^{90}Sr^{90}Y$ | $^{60}Co$ | $^{239}Pu$ | $^{239}Pu$ |
|---|---|---|---|---|---|---|
| Type of decay | e-capture | β⁻ | β⁻ | β⁻ | α | α |
| What is registered | γ | β | β | β, γ | α | α |
| Detector | NaJ(Tl) | СБМ-12 | СТС-5 | СБТ-7 | ДКПс-25 | ДКПс-25 |
| Count rate, counts/c | 1730 | 538 | 104 | 9.7 | 46.8 | 44.3 |
| Averaging time, hour | 4 | 17 | 22 | 23 | 19 | 20 |
| Standard deviations, % | 0.02 | 0.025 | 0.036 | 0.11 | 0.08 | 0.08 |
| Max. drop from December 12 till December 22, % | 0.25 | 0.051 | 0.054 | 0.15 | 0.16 | 0.15 |

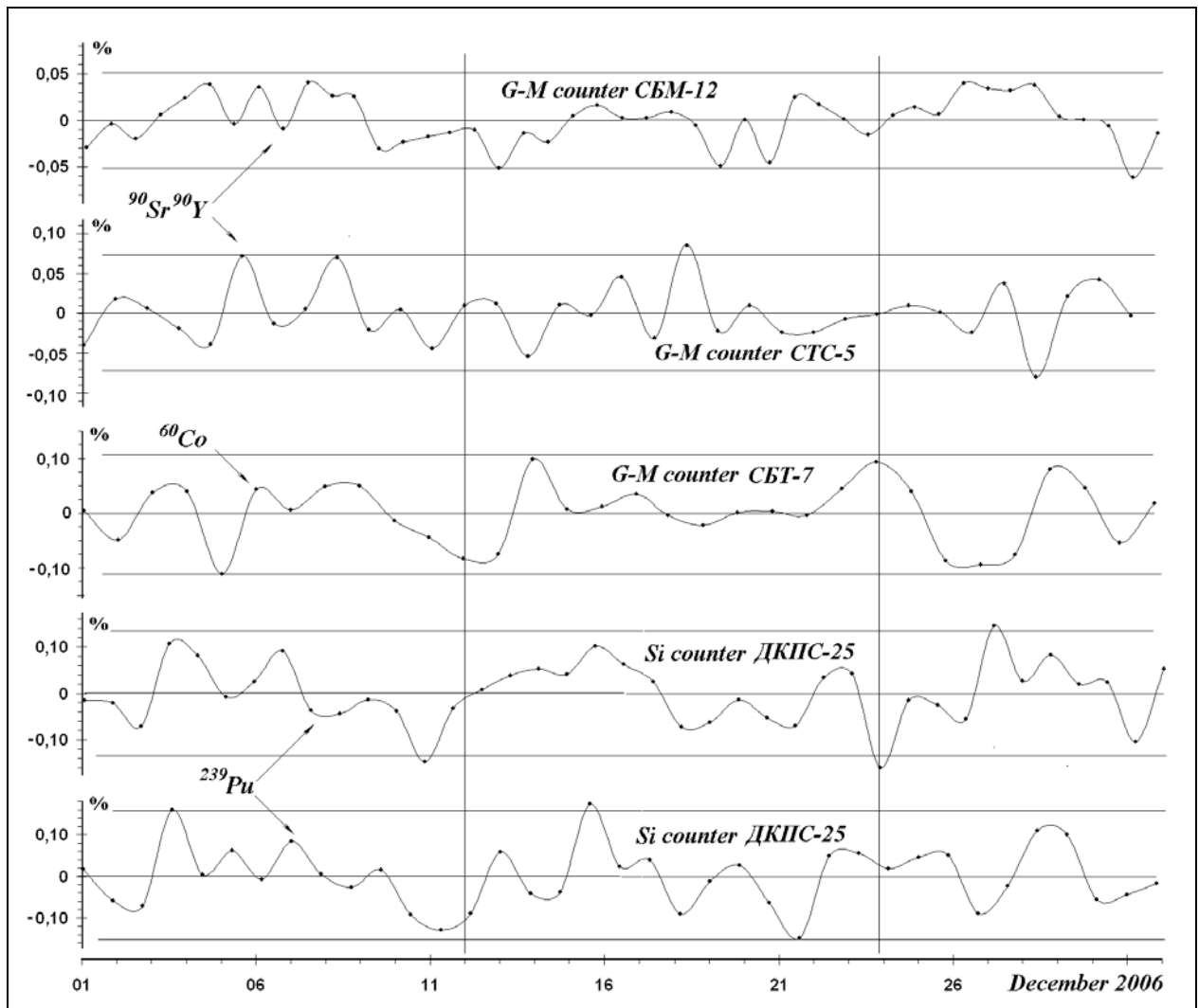

**Fig. 2.** Deviation from the average count rates for β decays of $^{90}Sr^{90}Y$ and $^{60}Co$ sources and for α decays of $^{239}Pu$ source (%). Horizontal lines correspond to the average values and levels of two standard deviations from the averages. Between verticals - time of the anomalous drop of $^{54}Mn$ count rate (see fig.1).

It is visible from table 1 and fig. 2 that during the anomalous decrease in count rate of $^{54}Mn$ between 12 and 22 December 2006, the count rates for both β and α sources in our installation did not drop below the value of 2 standard deviations from the average. This implies absence of statistically significant changes in the count rates about 0.1%.

**Measurement results from October-November 2003.**
In October and November 2003, several very powerful solar flares were registered [5]. It would be of great interest to analyze measurements of the count rates of radioactive decays at the time, with the aim to find possible deviations from the normal course for this process. The outcomes of count rate measurements of $^{90}Sr\ ^{90}Y$ and $^{60}Co$ β sources, and $^{239}Pu$ α source are shown in the Table 2 and Fig.3. In difference from described above measurements in December 2006, for $^{239}Pu$ radioactivity measurements the G-M counter СБТ-11 with a window from thin mica passing α particles was used. Such counter, in difference from the silicon detector, apart from α particles, registers β and γ radiation also.

**Table 2.** Sources, detectors and measurements in October -November 2003.

| Radionuclides | $^{90}Sr^{90}Y$ | $^{90}Sr^{90}Y$ | $^{60}Co$ | $^{239}Pu$ |
|---|---|---|---|---|
| Type of decay | β⁻ | β⁻ | β⁻ | α |
| What is registered | β | β | β, γ | α, β, γ |
| Detector | СБМ-12 | СТС-5 | СБТ-7 | СБТ-11 |
| Count rate, counts/c | 576 | 114 | 14.5 | 48.2 |
| Averaging time, hour | 6.3 | 4 | 16 | 5.6 |
| Standard deviations, % | 0.035 | 0.08 | 0.11 | 0.12 |
| Max. drop after flares , % | 0.07 | 0. 21 | 0.19 | 0.26 |

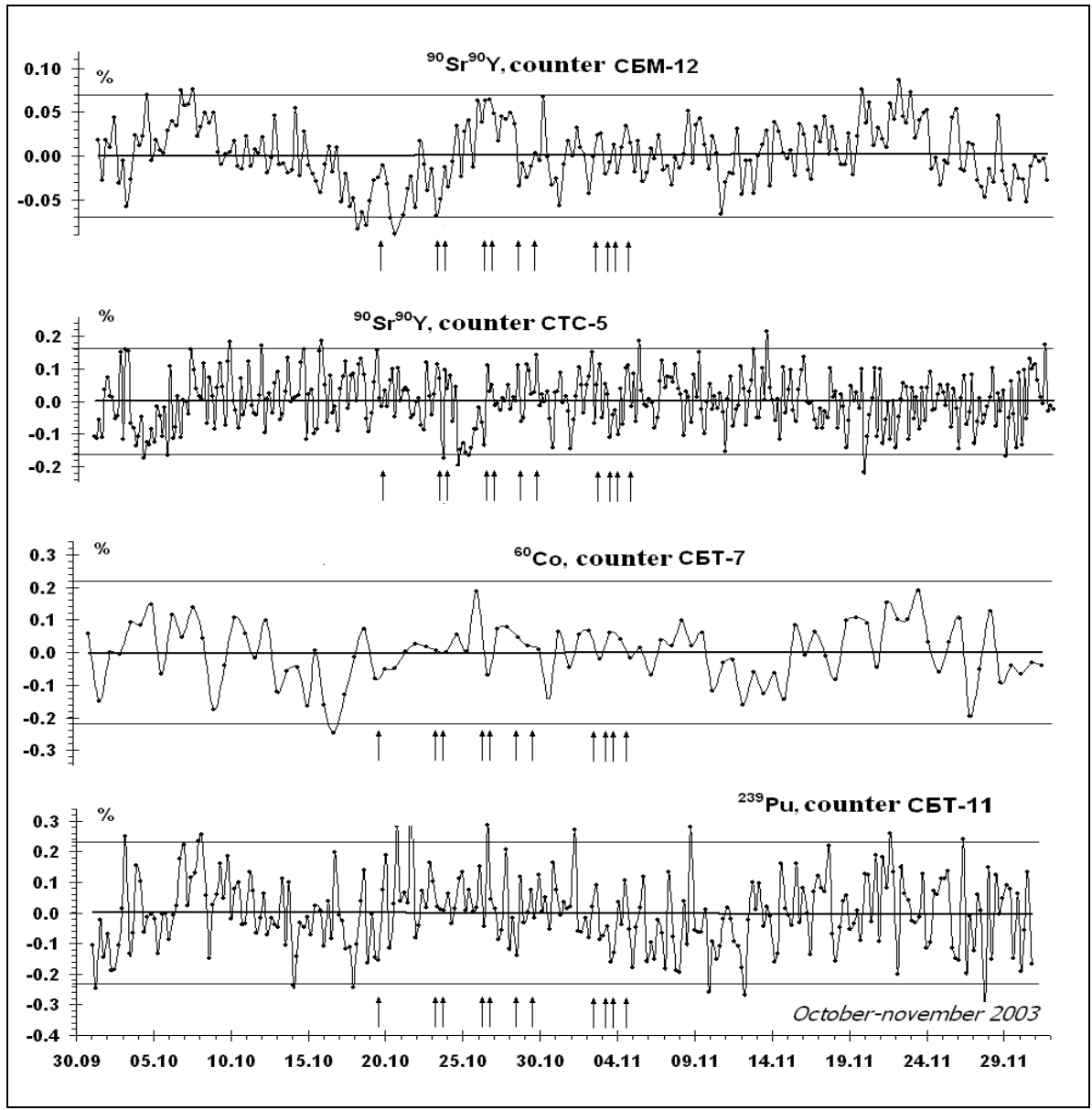

**Fig. 3.** Deviations from the average count rates of $^{90}Sr^{90}Y$, $^{60}Co$ and $^{239}Pu$ sources (%). Horizontal lines correspond to the average values and levels of two standard deviations from the averages. The arrows indicate moments of the most powerful solar flares.

Some of the measurements results differ from average by more than two standard deviations. However, the number of such deviations does not exceed the threshold allowed by the statistics of random events. Therefore, it is possible to state, that the series of powerful solar flares between 19 October and 4 November did not affect noticeably the activity of β sources $^{90}Sr^{90}Y$ and $^{60}Co$ and of α source $^{239}Pu$.

**Anomalous decreases in radioactivity distanced from solar flares**

So, our measurements have not registered any reliable drops in the count rates of α and β decays near the time of high-power solar flares. Nonetheless, there have been two cases of apparent deviations from the normal course of radioactive decays during the 12 years of the functioning of our installation. The most spectacular event of such kind was registered on 19 and 20 June 2004 (Fig.4).

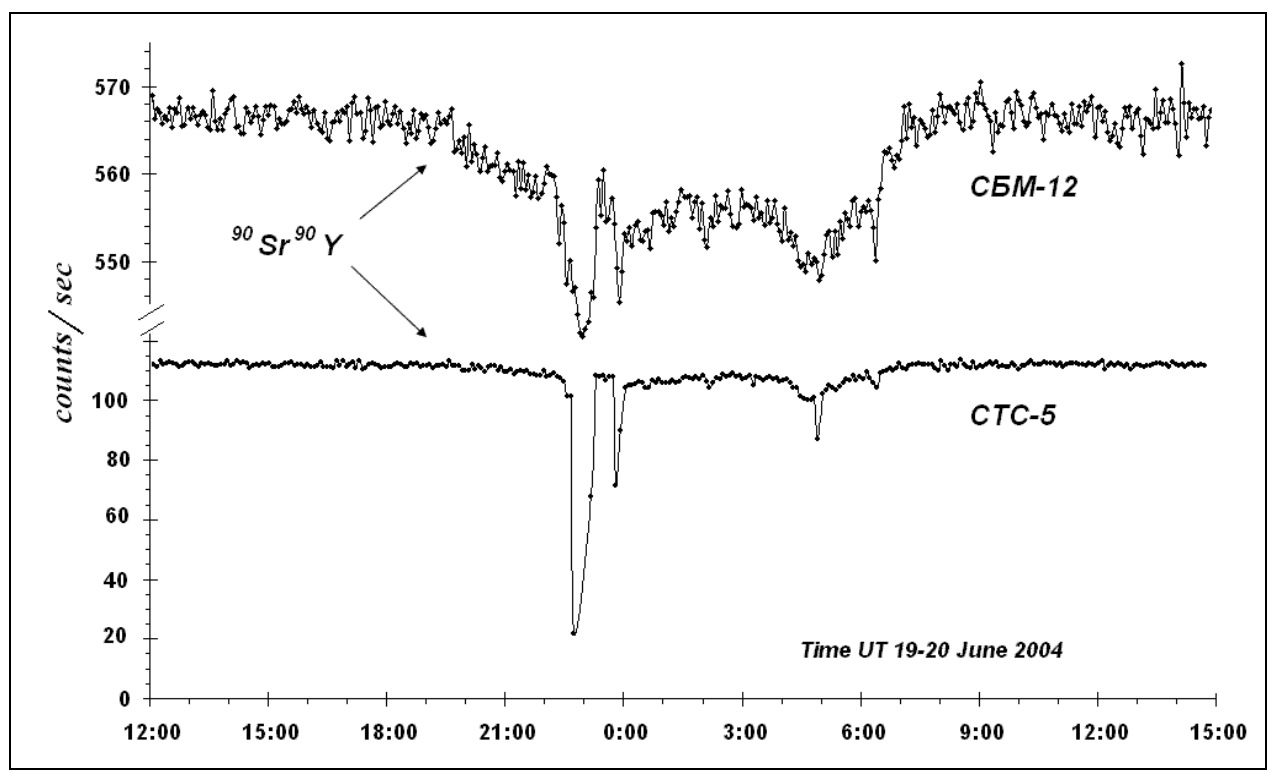

**Fig. 4**. Anomalous behaviour of count rate for $^{90}Sr$-$^{90}Y$ β source on 19 and 20 June 2004

Two different G-M counters (СБМ-12 and СТС-5) registered particles from one $^{90}Sr$-$^{90}Y$ β source. The first counter reacted to the particles almost the whole β-spectrum of both radionuclides, the other was sensitive only to the particles from $^{90}Y$ with energies close to the high end of the spectrum. The anomalous behaviour of signals lasted from 19:30 UTC, 19 June to 07:20, 20 June 2004. For the channel of first counter, a drop of up to 5% was observed, whereas for the second counter more than five-multiple drop of the count rate took place. No other similar event has been registered in these channels for all the 10-year period of observations. In other channels of multichannel installation recording count rate from a few α and β sources, radiation background, noise of transistors, frequency of quartz generators, temperature around the installation, anything exotic had not taken place at this time.

The registered effect is so great that looks improbable. The supposition arises that this anomaly was caused by inaccuracy of recording equipment or by great modifications in an environment. But the records of course of temperature, atmospheric pressure and air humidity show a lack of any anomalies at this time. Moreover, such drop of count rate cannot arise from modifications of environment in reasonable limits. Such decrease of count rate can take place at G-M counters feeding voltage reduction from 400 V to 200-250 V. But this voltage was monitored, and its variations during anomaly did not exceed 1V. Therefore it is necessary to recognize possible, that the large drop of count rate is connected just to decrease in the radioactivity itself.

One more statistically reliable drop of count rate took place between 10:00 UTC, 10 May and 04:00, 12 May 2002 (source $^{60}Co$ with detector СБМ-12). The diminution of count rate reached 6 % (Fig.5). No other similar anomalies have been detected during more than 10 years of the measurements of α and β radioactivity with our installation.

It is important to note, that neither in May 2002, nor in June 2004, when the anomalous drops of count rates were recorded by our installation, powerful solar flares were not registered [5].

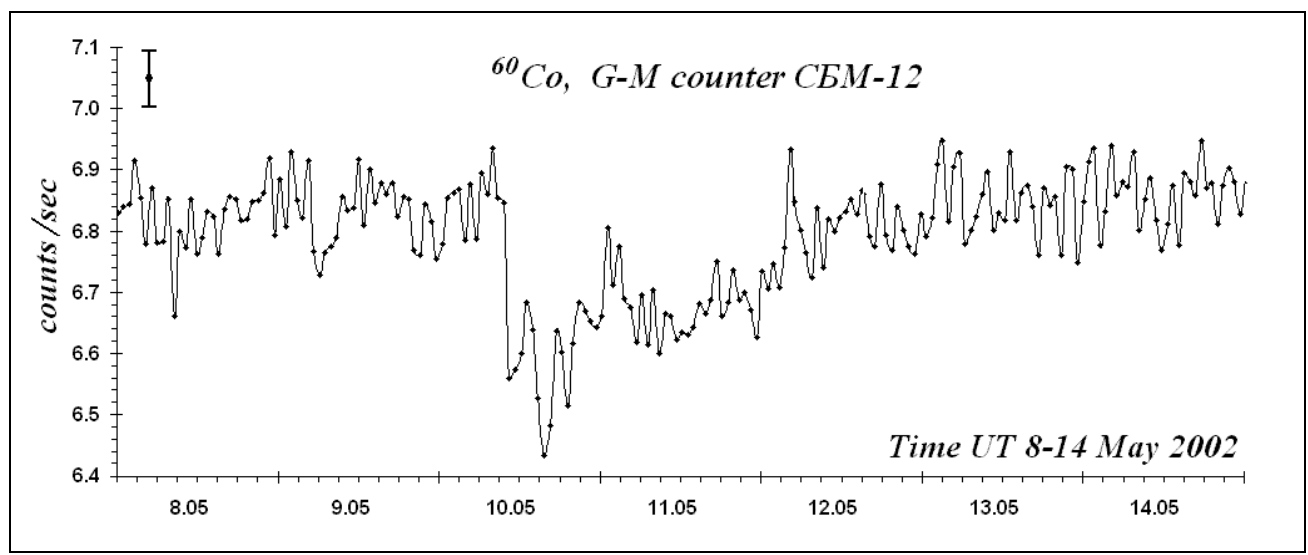

**Fig. 5**. Anomalous behaviour of the count rate from $^{60}Co$ source during 9-12 May 2002

**Conclusions**

The obtained experimental testimonies on effect of an anomalous radioactivity decrease are too poor to make unequivocal conclusions. However, they are sufficient for the preliminary judgment about some properties of this appearance, if it really exists.

1. It is rare occurrences. Summarized duration of radioactivity measurements of $^{90}Sr^{90}Y$, $^{60}Co$ and $^{40}K$ β sources on our installation is about 30 years. For this time only two cases statistically reliable drop of a radioactivity by a common duration about two days are registered. By dividing time of observation of the effect on common time of radioactivity measurements, it is possible to estimate probability of a detection of the effect: about $2 \cdot 10^{-4}$. It should be noted that the low activity of the used sources does not allow to conclude presence or lack of the effect with magnitudes less than 0.1 % from average value.

2. Summarized duration of α radioactivity measurements ($^{239}Pu$, $^{235}U$, $^{238}U$) is about 20 years. For this time, the effect of an anomalous drop of α radioactivity is not registered. Hence, there is no evidence to judge presence or lack of this effect in alpha radioactivity now.

3. In May 2002 9 various α and β sources, in June 2004 8 sources were measured simultaneously on our installation. In both cases, the effect was exhibited only at one from some sources. It testifies that the reason, calling this appearance, acts on different sources differently, even if they are located closely and the emitters contain identical radionuclides.
4. The absence of the effect of an anomalous decrease in the radioactivity near the times of strong solar flares and its presence at the time when no significant solar flares were registered, show that the idea [1,2] about connection of considered effect with solar flares is doubt. Most likely, the proximity in timing of powerful solar flares and drop in count rate of $^{54}Mn$ in December 2006 is casual coincidence. But it is also likely, that it is a result of some specific features of the electron capture. It is known that the electronic capture is distinguished among other types of β decays by some sensitivity to exterior actions via influence to electronic envelopes of atoms [6,7]. It seems reasonable to assume that the nuclei which are capable to electronic capture could be more sensitive to the agent calling an anomalous drop of a radioactivity.

Extraordinarity of the discovered phenomenon makes a problem of the further researches with the purpose of its reliable confirmation or refutation very important. This task requires organization of long high-precision radioactivity measurements for greater number of radionuclides. It is possible, that the effect of an anomalous drop in radioactivity can be revealed by a thorough analysis of the data gathered for other purposes, for example during half-lives measurements of long-lived radionuclides.